\documentclass[aps,prl,twocolumn,superscriptaddress,showpacs]{revtex4}

\usepackage{graphicx}

\begin{document}

\title{Determination of the intrinsic anomalous Hall effect of SrRuO$_3$}

\author{R. Mathieu\cite{cryo}}

\affiliation{Spin Superstructure Project (ERATO-SSS), JST, AIST Central 4, Tsukuba 305-8562, Japan}

\author{C. U. Jung}

\affiliation{Spin Superstructure Project (ERATO-SSS), JST, AIST Central 4, Tsukuba 305-8562, Japan}

\author{H. Yamada}
\affiliation{Correlated Electron Research Center (CERC), AIST Central 4, Tsukuba 305-8562, Japan}

\author{A. Asamitsu}
\affiliation{Spin Superstructure Project (ERATO-SSS), JST, AIST Central 4, Tsukuba 305-8562, Japan}
\affiliation{Cryogenic Research Center (CRC), University of Tokyo, Bunkyo-ku, Tokyo 113-0032, Japan}

\author{M. Kawasaki}
\affiliation{Correlated Electron Research Center (CERC), AIST Central 4, Tsukuba 305-8562, Japan}
\affiliation{Institute for Materials Research, Tohoku University, Sendai 980-8577, Japan}

\author{Y. Tokura}
\affiliation{Spin Superstructure Project (ERATO-SSS), JST, AIST  Central 4, Tsukuba 305-8562, Japan}
\affiliation{Correlated Electron Research Center (CERC), AIST Central 4, Tsukuba 305-8562,
Japan}
\affiliation{Department of Applied Physics, University of Tokyo, Tokyo 113-8656, Japan}

\begin{abstract}

The anomalous Hall effect (AHE) of epitaxial SrRuO$_3$ films with varying lattice parameters is investigated, and analyzed according to the Berry-phase scenario. SrRuO$_3$ thin films were deposited on SrTiO$_3$ substrates directly, or using intermediate buffer layers, in order to finely control the epitaxial strain. The AHE of the different films exhibits intrinsic features such as the sign change of the Hall resistivity with the temperature, even for small thicknesses of SrRuO$_3$. However, the anomalous Hall conductivity is greatly reduced from its intrinsic value as the carrier scattering is increased when the epitaxial strain is released. We argue that the AHE of fully strained SrRuO$_3$ film with low residual resistivity represents the intrinsic AHE of SrRuO$_3$.

\end{abstract}


\maketitle

The intrinsic nature of the anomalous Hall effect of itinerant ferromagnets\cite{kl} has recently been demonstrated theoretically\cite{NJPSJ,Fang}. This geometrical origin is related to Berry-phase mechanisms associated with the local spin texture of the material. This was thus first evidenced experimentally in ferromagnets with noncoplanar spin structure\cite{pyro,Jungwirth}. Interestingly, the topological origin of the AHE is also perceived in ordinary ferromagnets, like bcc Fe\cite{MacDonald}, or the well known itinerant ferromagnet SrRuO$_3$ (SRO), with essentially collinear magnetic moments\cite{Fang,SCRO}. While the extrinsic models of the AHE predict a Hall resistivity $\rho_{xy}$ proportional to the magnetization\cite{note,EX}, the intrinsic picture implies that the Hall conductivity $\sigma_{xy}$ critically depends on the details of the band structure of the material, and particularly on band crossing points acting as ``magnetic monopoles'' in the momentum space\cite{Fang}. $\sigma_{xy}$ is thus in the latter case very sensitive to the chemical potential and spin polarization of the material.\\
\indent In agreement with these predictions, $\sigma_{xy}$ of Ca doped SRO was experimentally found to scale fairly well with the magnetization\cite{SCRO}. The ferromagnetic interaction and associated Curie temperature $T_c$ decrease with increasing Ca content\cite{referee}. As a result, the $\rho_{xy}$($T$) curve of the substituted films is globally shifted to lower temperatures as the Ca content increases, retaining its characteristic shape. However the replacement of Sr by Ca introduces a large disorder, resulting in a fairly large residual resistivity at low temperatures for the films with Ca contents larger than 10\% \cite{SCRO}. In order to control the carrier scattering and $T_c$ more finely, we have grown thin films of SrRuO$_3$ using different buffer layers, on SrTiO$_3$ (STO) substrates with different crystallographic orientations. The high quality of the thin films epitaxially grown on STO, i.e grown without buffer, is demonstrated. The in-plane strain, originally compressive, is gradually minimized using buffer layers, and turned into tensile, yielding the increase of $T_c$ toward the bulk value, as well as the increase of the residual resistivity. The strain is minimum in a SRO film grown on a CaHfO$_3$ buffer layer. The AHE of this film is compared to those of bulk SRO single crystal and epitaxially strained SRO film, as well as a SRO/STO superlattice and an ultrathin SRO film for comparison. While the AHE of the strained films is essentially intrinsic in origin, we observe that $\sigma_{xy}$ is largely reduced from its intrinsic value as the electron scattering increases.

Thin ($\sim$ 150 monolayers (ML), $\sim$ 600 {\AA}) films of SrRuO$_3$ were epitaxially grown on high quality SrTiO$_3$ single-crystal substrates by pulsed laser deposition (PLD). The SRO films were either grown directly on the (001) and (110) surfaces of STO (the resulting films being referred to as STO 001 and STO 110 respectively), or using 600 {\AA} thick buffer layers of CaHfO$_3$ (CHO) or Ca$_{0.8}$Sr$_{0.2}$SnO$_3$ (CSSO)\cite{JK} (CHO/STO 001, CHO/STO 110, and CSSO/STO 110). The detailed structural characterization of the thin films was performed by x-ray diffraction on a four-circle diffractometer with Cu K$\alpha$ source (K$\alpha_1$ and K$\alpha_2$ beams)\cite{Taka}. The thickness of some films was confirmed using scanning electron microscopy (SEM). A ultrathin SRO film (5 ML), $\sim$ 20 {\AA}, as well as a (SRO(5 ML)/STO(10 ML))$_{\rm x10}$ superlattice (SL) were grown for comparison on STO (001). A bulk single crystal SrRuO$_3$ was also prepared using a flux method. The magnetization data was recorded on a MPMSXL SQUID magnetometer using a magnetic field applied normal to the plane of the films. The films were then patterned in a six-lead Hall bar geometry using conventional photo-lithography and Ar ion etching for transport measurements. The Hall resistivity  $\rho_{H}$ was measured with a PPMS6000 system together with the longitudinal resistivity $\rho_{xx}$ as a function of $H$ and $T$. The anomalous  resistivity $\rho_{xy}$  was extrapolated to $H$ = 0 from $\rho_{H}$  vs $H$ measurements up to $H$ = $\pm$ 5 or 9 T at constant temperatures (from 2 K to 240 K) after subtraction of the ordinary Hall contribution,  and the transverse conductivity $\sigma_{xy}$ was estimated as -$\rho_{xy}/\rho^2_{xx}$. A small (as the patterned leads are nearly symmetric) magnetoresistance was removed by subtracting $\rho_{H}$($-H$) to $\rho_{H}$($H$). 

\begin{figure}[h]
\includegraphics[width=0.46\textwidth]{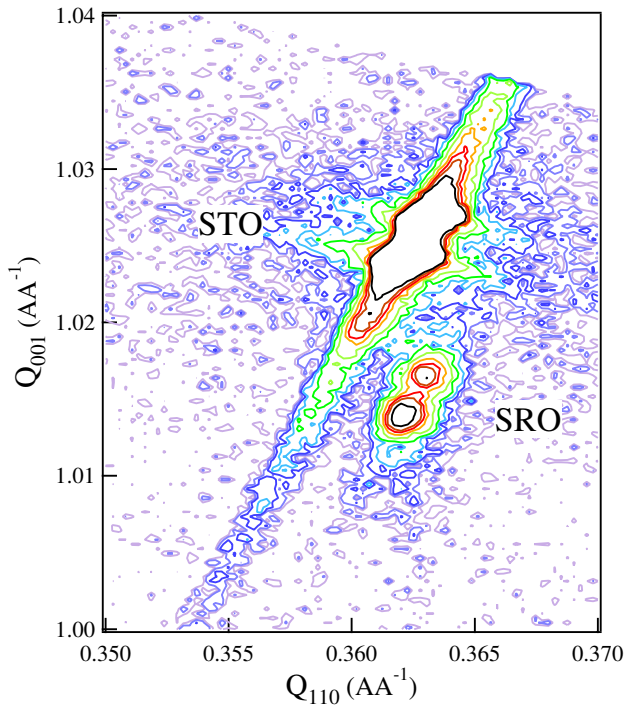}
\caption{(color online) Reciprocal space mapping for the (114) peaks of SRO and STO in the STO 001 film. The x-ray radiation consists of K$\alpha_1$ and K$\alpha_2$ beams, which yields the splitting of the (114) peaks.}
\label{fig:xray}
\end{figure}

\begin{figure}[h]
\includegraphics[width=0.46\textwidth]{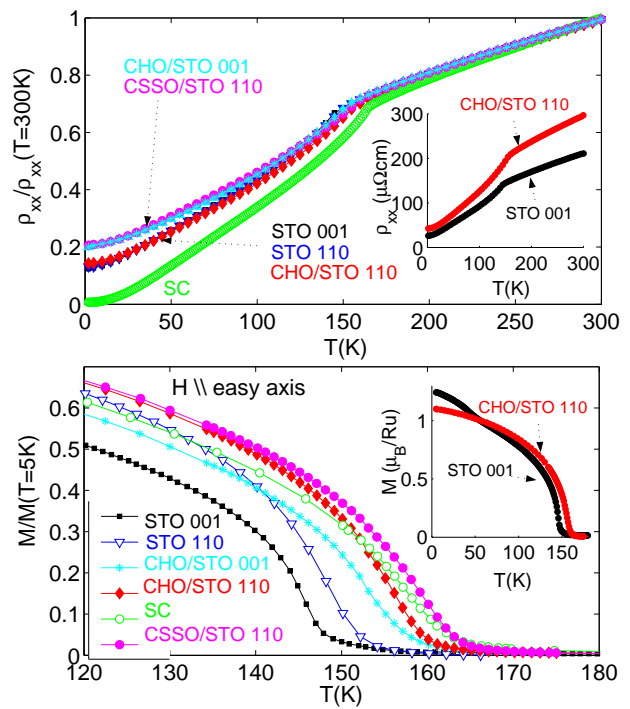}
\caption{(color online) Temperature dependence of the zero-field resistivity $\rho_{xx}$ (upper panel) and normalized magnetization $M$ (lower panel). The magnetization is recorded in $H$ = 500 Oe on re-heating after cooling in $\mu_0H$ = 7T, applied along the easy axis of magnetization. The single crystal data is added for comparison. The insets shows the temperature dependence of $\rho$ and $M$ for STO 001 and CHO/STO 110. Ruthenates are very robust against oxygen deficiency, and the analysis of the lattice parameters of the films indicate no or insignificant Ru deficiency. }
\label{fig:rmt}
\end{figure}

While the synthesis of clean bulk single-crystals is relatively difficult, high quality thin films of SRO can be grown by PLD on STO (001) substrates. While SRO has an orthorhombic structure at room temperature, it adopts a cubic structure at the PLD growth temperature ($\sim$ 1000 K). During the cooling subsequent to the deposition, the SRO layers undergo a structural phase transition, and their final structure will be affected by the epitaxial strain from the substrate\cite{Maria}. Our high quality STO (001) substrates have only 1-unit-cell-high steps and 300 nm-wide terraces with atomically well-defined surface\cite{Kawa}. Such a 'step and terrace' substrate enhances the effect of the epitaxial strain, resulting in the tetragonal structure without twin even on (001) substrates\cite{Kawa}. As an illustration, the x-ray diffraction data of the STO 001 film is shown in Fig.~\ref{fig:xray}. As seen in this figure, there is a single (114) peak for SRO (splitted because two different radiations K$\alpha_1$ and K$\alpha_2$ are employed), which appears at the same value on the horizontal axis as the (114) peak of the STO substrate. There is thus no orthorhombic/rhombohedral distortion nor twinning, and the tetragonal lattice of SRO has in-plane lattice constants identical to those of its substrate\cite{Taka,SCRO}. 

\begin{figure}[ht]
\includegraphics[width=0.48\textwidth]{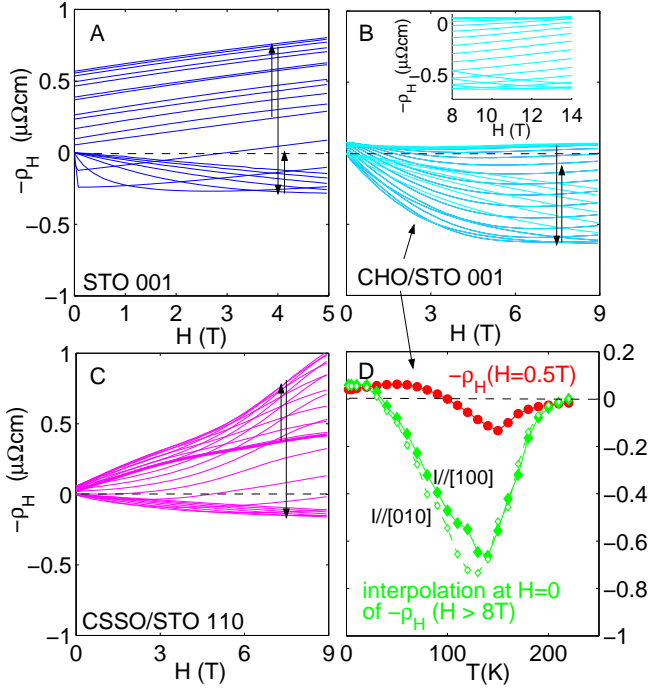}
\caption{(color online) Panels A, B, C: Magnetic field dependence of $\rho_H$ from 2K to 240K, measured every 10K; the arrows indicate increasing temperatures. The inset in the panel B shows $\rho_H$($H$) up to larger magnetic fields. The panel D shows the temperature dependence of $\rho_{xy}$ estimated either as $-\rho_{H}$($H$=0.5T) or from the high-field -$\rho_{H}$($H$) data. Two pieces of the CHO/STO 001 were patterned along different crystallographic directions; only marginal differences were observed in $\rho_{H}$ when feeding the current $I$ along the [001] and [010] directions. }
\label{fig:rhoh}
\end{figure}

The magnitude and sign of the epitaxial strain on SRO can thus be controlled by depositing buffer layers on STO prior to SRO. The temperature dependence of the normalized electrical resistivity and magnetization of the different SRO films described above is shown in Fig.~\ref{fig:rmt}. An inflection is observed in the resistivity curves of all samples near $T_c$ ($\sim$ 150 - 165 K). The different curves coincide above $T_c$, and depart from each other at low temperatures. The residual resistivity of the bulk single crystal is very low ($\sim$ 2 $\mu\Omega$cm). It increases by about an order of magnitude in the thin films. The SRO films deposited directly on our STO (001) are, as mentioned above, clamped by the substrate. The growth is coherent throughout the film, which adopts a tetragonal structure\cite{SCRO}. Due to the spin-orbit interaction, the easy axis of magnetization follows the elongation direction, i.e. the normal to the film plane\cite{JK,Izumi}. 

As seen in Fig.~\ref{fig:rmt}, the residual resistivity ratio [$RRR=\rho$(300K)/$\rho$(2K)] is relatively large for the STO 001 film, reflecting the coherent growth of the SRO layers. $T_c$ is, however, nearly 15 K lower than that of bulk single-crystalline SRO due to the large compressive strain\cite{CBEOM}. Compressive or tensile strains are likely to affect the distortion of the RuO$_6$ octahedra, and thus the Ru 4$d$ band spawning the ferromagnetic interaction\cite{newkawa}. $T_c$ can be increased by releasing the substrate strain, using intermediate buffer layers\cite{Terai,JK}, or simply removing the substrate\cite{CBEOM}. In the present case, the in-plane strain gradually changes from compressive (STO 001, STO 110) to tensile (CHO/STO 001, CSSO/STO 110)\cite{JK,Terai}. In the latter case, the relaxation of the epitaxial strain yield both the increase of $T_c$ and the decrease of the RRR, as the growth is not so coherent as in the case of STO 001 (see Fig.~\ref{fig:rmt}). The compressive epitaxial strain becomes minimum for CHO/STO 110, in which the SRO layer has lattice parameters similar to those of a bulk single crystal of SrRuO$_3$. However, the growth is still coherent, and the easy axis of magnetization remains perpendicular to the film plane. When the strain becomes tensile however, the easy axis of magnetization rotates toward the film plane direction, lying in the direction determined by the structure of the relaxed SRO layers.
\begin{figure}[h]
\includegraphics[width=0.46\textwidth]{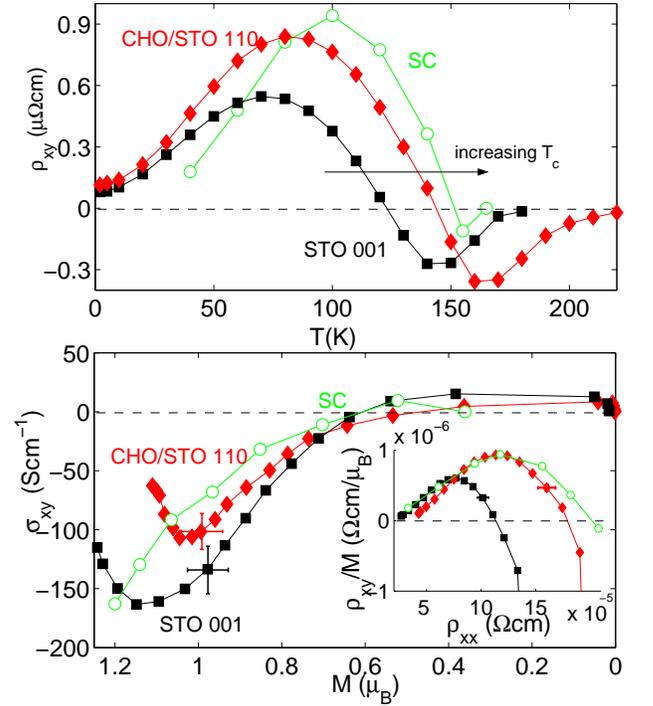}
\caption{(color online) Upper panel: Temperature dependence of the anomalous Hall resistivity $\rho_{xy}$ for STO 001 (largest compressive strain, filled squares), CHO/STO 110 (minimum strain, filled diamonds), and the bulk single crystal of SRO (SC, open circles).  Lower panel: The anomalous Hall conductivity $\sigma_{xy}$ is plotted as a function of the magnetization $M$. The inset shows the corresponding $\rho_{xy}/M$ ($\propto$ the extrinsic $R_s$\cite{note}) data plotted as a function of $\rho_{xx}$. Typical errorbars are added (see Ref. \cite{SCRO})}
\label{fig:rhoxy}
\end{figure}

This is illustrated in Fig.~\ref{fig:rhoh}, which shows the magnetic field dependence at different temperatures of the measured $\rho_H$. We here determine $\rho_{xy}$ by extrapolation to $H$ = 0 of the -$\rho_{H}$($H$) data. It is also common to define $\rho_{xy}$ as the value of -$\rho_{H}$ in a small magnetic field for which most of the magnetization change is completed, as for example -$\rho_{H}$($H$ = 0.5 T)\cite{pyro}. For the SRO films with the out-of-plane easy axis of magnetization (STO 001, STO 110, CHO/STO 110, the ultrathin SRO film, and the SRO/STO SL), both descriptions yield marginally different results (as exemplified in Fig.~\ref{fig:rhoh} A for STO 001), as  $\rho_H$ is linear with $H$ above 0.5 T. The ordinary Hall effect ($\sim$ -$R_oH$, with $R_o<$0) beeing responsible for the linear increase of -$\rho_H$($H$) in large magnetic fields. For the relaxed films, with the easy axis of magnetization nearly in plane (as CHO/STO 001 and CSSO/STO 110), the two methods yield quite different results (see Fig.~\ref{fig:rhoh} D). Measurements up to 14 T were performed to confirm the linearity of the high field data (as seen in Fig.~\ref{fig:rhoh} B). In the case of the CSSO/STO 110 film, $\rho_{H}$ is not linear with $H$ even for the largest magnetic fields, so that the $H$ = 0 extrapolation is not possible. The magnetic anisotropy field may thus be comparable or larger than 10 T in the relaxed films, making the estimate of $\rho_{xy}$ difficult as the ordinary and anomalous Hall contributions cannot be distinguished.

\begin{figure}[h]
\includegraphics[width=0.46\textwidth]{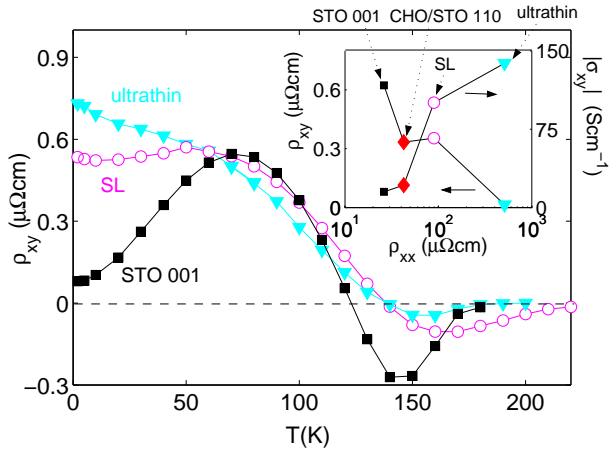}
\caption{(color online) Temperature dependence of the Hall resistivity $\rho_{xy}$ for STO 001, the ultrathin SRO film, and the SRO/STO superlattice (SL). The inset shows $\rho_{xy}$ and $\sigma_{xy}$ plotted as a function of $\rho_{xx}$ for $T$ = 2 K for these films and CHO/STO 110.}
\label{fig:rhoxyall}
\end{figure}

For the films with out-of-plane magnetization, the $\rho_{xy}$ estimated at different temperatures is plotted in the upper panel of Fig.~\ref{fig:rhoxy}. As observed in the Ca doped SRO films\cite{SCRO}, the different $\rho_{xy}$($T$) curves are very similar to each other, albeit shifted in temperatures according to the $T_c$ value of the sample. As shown in the lower panel of Fig.~\ref{fig:rhoxy}, $\sigma_{xy}$ scales fairly well with the magnetization, as predicted by the Berry phase scenario\cite{Fang,SCRO} (the $\sigma_{xy}(M)$ curve being characteristic of SrRuO$_3$). Since the samples have different resistivities and thus different degree of electronic scattering, slight variations of $\sigma_{xy}$ can be expected (see below). In the extrinsic picture, the $\rho_{xy}/M=R_s$ of a given material is a function of $\rho_{xx}$, and $T$\cite{note}. However, as seen in the inset of the lower panel of Fig.~\ref{fig:rhoxy}, the different  $\rho_{xy}/M$($\rho_{xx}$) curves of our SRO samples do not scale with each other in the measured range of temperatures. Furthermore, while $R_s$ is usually considered to be a monotonous (and little varying) function of the temperature, a sign change is observed as the temperature increases.

Remarkably, STO 001 films remain ferromagnetic down to very small thicknesses ($\sim$ 4 ML). As the substrate clamping is still active, the easy axis of magnetization remains out-of-plane. As seen in the main frame of Fig.~\ref{fig:rhoxyall}, the $\rho_{xy}$($T$) curves of an 5 ML SRO film and a SRO(5 ML)/STO(10 ML) SL show the ``intrinsic'' sign change below $T_c$. However, $\rho_{xy}$ and  $\rho_{xx}$ are greatly increased at low temperatures. The increase of $\rho_{xy}$ at low temperature may essentially be related to the large scattering (with scattering rate $\sim \sigma_{xx}^{-1}$ $\sim$ $\rho_{xx}$ ), as seen in the inset of Fig.~\ref{fig:rhoxyall}. Thus, as $\rho_{xx}$ increases, $\sigma_{xy}$ is greatly reduced from its intrinsic value, which can thus be estimated as $\sim$ 120 Scm$^{-1}$, as obtained for the fully strained SRO film (STO 001) a low temperatures. Interestingly, the intrinsic AHE of SRO can only be estimated in thin films with moderate disorder, such as STO 001. If the so-called clean limit was to be reached, e.g. in thin films with extremely low resistivity, the measured $\sigma_{xy}$ would rapidly depart from its intrinsic value\cite{Onoda}.\\

To summarize, we have investigated the AHE of SrRuO$_3$ thin films with different crystallographic structures and thicknesses. We have observed that while the in-plane strain can be minimized, yielding the increase of $T_c$ toward the bulk single crystal value, the AHE is affected by the additional carrier scattering brought forth by the release of the strain. Nevertheless, intrinsic features are still observed for all the films, yielding the scaling of $\sigma_{xy}$ with the magnetization as predicted by the Berry-phase scenario. The measured $\sigma_{xy}$ is found to be greatly reduced from its intrinsic value as the scattering rate increases. The intrinsic AHE of SRO, which is essentially scattering-rate independent, can thus experimentally be estimated only in a limited range of longitudinal resistivities.\\

\indent We thank Drs. S. Onoda, M. Onoda, and Prof. N. Nagaosa for enlightening discussions.

\end{document}